\documentclass[12pt,intlimits,reqno]{amsart}
\usepackage{enumerate}
\usepackage{amssymb}
\usepackage{amsthm}
\usepackage{mathrsfs}
\usepackage[OT4]{fontenc}
\usepackage{verbatim}
\usepackage{antpolt}
\usepackage{enumerate}
\usepackage{bbm}

\newtheorem{thm}{Theorem}[section]

\newtheorem{prop}[thm]{Proposition}
\theoremstyle{definition}

\theoremstyle{remark}

\numberwithin{equation}{section}

\newcommand{\pair}[2]{\langle #1\;;\;#2 \rangle}
\newcommand{\abs}[1]{\left\vert#1\right\vert}
\newcommand{\set}[1]{\left\{#1\right\}}
\newcommand{\R}{\mathbb R}

\newcommand{\No}{\mathbb{N}\cup\{0\}}

\newcommand{\Z}{\mathbb Z}
\newcommand{\C}{\mathbb C}
\newcommand{\N}{\mathbb N}

\renewcommand{\H}{\mathcal{H}}
\renewcommand{\Re}{\mathrm{Re}}

\renewcommand{\u}{\mathfrak{u}}

\newcommand{\be}{\begin{equation}}
\newcommand{\ee}{\end{equation}}
\newcommand{\ben}{\begin{enumerate}}
\newcommand{\een}{\end{enumerate}}
\newcommand{\prf}[1]{\begin{proof}#1\end{proof}}
\DeclareMathOperator{\const}{const}

\newcommand{\pb}{\{\,\cdot\,,\,\cdot\,\}}

\DeclareMathOperator{\Tr}{Tr}
\begin{document}

\title[Hierarchy of integrable Hamiltonians describing$\ldots$]{Hierarchy of integrable Hamiltonians describing of nonlinear $n$-wave interaction}
\author{A. Odzijewicz, T. Goli\'nski}
\dedicatory{University in Bia{\l}ystok\\
Institute of Mathematics\\
Lipowa 41, 15-424 Bia{\l}ystok, Poland\\
email: aodzijew@uwb.edu.pl, tomaszg@alpha.uwb.edu.pl\\
}

\begin{abstract}
In the paper we construct an hierarchy of integrable Hamiltonian systems which describe the variation of $n$-wave envelopes in nonlinear dielectric medium. The exact solutions for some special Hamiltonians are given in terms of elliptic functions of the first kind.
\end{abstract}

\maketitle

\section{Introduction}

The describtion of the nonlinear $n$-wave interaction plays important role in many areas of physics including optics, accoustic, plasma and fluid physics, see e.g. \cite{holm1,david-holm,alber-marsden,kaup}. For example one can find the solutions of three-wave equations in early works in the area of nonlinear optics \cite{armstrong}.

In this paper we study a hierarchy of integrable Hamiltonian systems on the
space of linear maps $L(\H_-,\H_+)$ between two complex finite dimensional Hilbert spaces
$\H_-$ and $\H_+$ which is the particular case of a more general hierarchy defined on the Banach Lie--Poisson space related to the restricted Grassmannian, see \cite{GO-grass}. This hierarchy, as we will show, can be used to describtion of the variation of plenary wave envelopes through the dielectric nonlinear medium.

In particular we study in detail the case when $\dim \H_+=2$ and $\dim \H_-=3$. The case when $\dim \H_+=2$ and $\dim \H_-=2$ is also investigated up to giving explicite formulas for solutions. Hamiltonians in these two cases describe  a nonlinear optical system consisting of six and four waves  interacting in a nonlinear medium displaying Kerr-like effects and causing convertion of light modes. 

For the paper self-sufficiency
we give in Section \ref{sec:physback} a short presentation of the wave optics background concerning the slow moving wave envelopes in the nonlinear medium.

In Section \ref{sec:hierarchy} the construction of the hierarchy of integrable Hamiltonian systems mentioned above is given and is shown that it has a rich family of integrals of motion in involution.


In Section \ref{sec:2+3} we investigate the $(2+3)$-dimensional system.  Applying reduction procedure we construct the angle-action coordinates for this systems and integrate it in quadratures.

The $(2+2)$-dimensional case we consider in Section \ref{sec:2+2}. We show that this case can be obtained as a reduction of the previous one and find solutions for it in terms of elliptic functions of the first kind.

To conclude the paper we shortly discuss in Section \ref{sec:phys} the physical interpretation of separate terms of one of the Hamiltonian in $(2+3)$-dimentional case.

\section{Wave propagation in dielectric media}\label{sec:physback}

The nonlinear wave optics deals with the interaction of the eletromagnetic waves with the medium. The complicated character of this interaction is manifested in the nonlinear dependence of the medium polarization field $\mathbf P(t,\mathbf x)$ on the electric field $\mathbf E(t,\mathbf x)$. Since the efects caused by the magnetic field are much weaker then the ones for which the electric field is responsible, so, they are usually neglected in wave optics problems, see e.g. \cite{Butcher-Cotter}.

Let us start from Maxwell equations in dielectric medium without the conduction current
\begin{align}\label{max}
 \nabla \times \mathbf E(t)&= - \frac{\partial}{\partial t} \mathbf B(t)\\
\nabla \cdot (\epsilon_0\mathbf E(t)+\mathbf P(t))&=0 \\
\nabla \times \mathbf H(t)&= \epsilon_0 \frac{\partial}{\partial t} \mathbf E(t)+ \frac{\partial}{\partial t} \mathbf P(t)\\
\label{max4} \nabla \cdot \mathbf B(t)&=0,
\end{align}
where $\frac{\partial\mathbf P}{\partial t}(t)$ is the polarization current. In optics one usually has deal with nonmagnetic media what is expressed in the dependence 
\be \label{magn}\mathbf B(t)=\mu_0\mathbf H(t)\ee
between the magnetic induction and the magnetic field. On the other hand the polarization $\mathbf P(t)$ dependence on the electric field has in general the nonlinear functional character
\be \label{polar}\mathbf P(t)=\mathbf P[\mathbf E(t)].\ee
Substituting \eqref{magn} into \eqref{max}-\eqref{max4} one obtains 
\be \label{e4}\nabla\times(\nabla\times \mathbf E(t))=-\frac1{c^2}\frac{\partial^2}{\partial t^2} \mathbf E(t)- \mu_0 \frac{\partial^2}{\partial t^2}\mathbf P(t).\ee 
Next, expressing $\mathbf E(t)$ and $\mathbf P(t)$ in terms of their Fourier transforms $\mathbf E(\omega)$ and $\mathbf P(\omega)$ one rewrites equation \eqref{e4} and \eqref{polar} as follows
\be \label{e5}\nabla\times\nabla\times\mathbf E(\omega)=\frac{\omega^2}{c^2}\mathbf E(\omega)+\omega^2\mu_0\mathbf P(\omega)\ee
\be \label{polar2}\mathbf P(\omega)=\mathbf P[\mathbf E(\omega)].\ee
The equation \eqref{polar2} in the most general setting has form
\be \mathbf P(\omega)=\sum_{n=0}^\infty \epsilon_0\int_{-\infty}^\infty d\omega_1\int \ldots\int d\omega_n\; \chi^{(n)}(\omega_1,\ldots,\omega_n)(\mathbf E(\omega_1),\ldots,\mathbf E(\omega_n)),\ee
e.g. see \cite{Butcher-Cotter}, where the $n$-linear form $\chi^{(n)}(\omega_1,\ldots,\omega_n)$, called $n^{th}$-susceptibility tensor, describes the order of nonlinearity of the interaction of the electric field with the medium.

The ones of possible solutions of \eqref{e5}-\eqref{polar2} are the running plane waves
\be \label{planewave}\mathbf E(\omega)=\hat {\mathbf E}(\omega,\kappa)e^{i\kappa}\ee
where $\kappa:=\mathbf k\cdot \mathbf r$ and $\mathbf k\cdot \hat{\mathbf E}(\omega,\kappa)=0$. Substituting \eqref{planewave} into \eqref{e5} we obtain
\be \label{e9}\frac{d^2}{d\kappa^2}\hat{\mathbf E}(\omega,\kappa)+2i\frac{d}{d\kappa}\hat{\mathbf E}(\omega,\kappa)=
\hat{\mathbf E}+\frac1{k^2}\mathbf P[\hat {\mathbf E}(\omega,\kappa)e^{i\kappa}]e^{-i\kappa}.\ee
If the variations of wave envelopes $\hat{\mathbf E}(\omega,\kappa)$ are sufficiently slow in variable $\kappa$ then we can assume 
\be \abs{\frac{d^2}{d\kappa^2}\hat{\mathbf E}(\omega,\kappa)}<<\abs{\frac{d}{d\kappa}\hat{\mathbf E}(\omega,\kappa)}\ee
and neglect the second derivative in \eqref{e9}.

So, in the slowly-varying approximation we have a system of first order differential equations
\be \label{e11}\frac{d}{d\kappa}\hat{\mathbf E}(\omega,\kappa)=\frac1{2i}
\hat{\mathbf E}+\frac1{2ik^2}\mathbf P[\hat{\mathbf E}(\omega,\kappa)e^{i\kappa}]e^{-i\kappa}\ee
on the infinite family of functions $\mathbf E(\omega,\kappa)$ parametrized by the real numbers $\omega\in\R$.

In practical applications one restricts usually its attention to the finite number of the running plane waves which are labeled by the frequencies $\omega_1,\ldots,\omega_N$ and polarizations of the envelopes $\hat{\mathbf E}(\omega_1,\kappa)$,$\ldots$,$\hat{\mathbf E}(\omega_N,\kappa)$. In this case the systsem of equations \eqref{e11} reduces to the finite system of Hamilton equations
\begin{align}
\frac{d}{d\kappa} z_i&=i\frac{\partial H}{\partial \bar z_i}\\ 
\frac{d}{d\kappa} \bar z_i&=-i\frac{\partial H}{\partial z_i}\\ 
\end{align}
with the Hamiltonian $H=H(z_1,\ldots,z_{2N},\bar z_1,\ldots z_{2N})$ which depends in nonlinear way on the independent modes $z_1,\ldots,z_{2N}$ describing a system of $N$ running plane waves which slowly vary in space and the evolution of a system of nonlinearly interacting harmonic oscillators \cite{kummer-n}.

In the next section we will construct a hierarchy of such type integrable Hamiltonian systems.

\section{Hierarchy of Hamiltonian integrable systems on $Mat_{M\times N}(\C)$}\label{sec:hierarchy}

In the paper \cite{GO-grass} we have constructed a hierarchy of Hamiltonian integrable systems on
$L^2(\H_-,\H_+)$, where $\H_-$ and $\H_+$ are complex separable Hilbert spaces (finite or infinite dimensional) and $L^2$ denotes the class of Hilbert--Schmidt operators. In this paper we will restrict our considerations to the subcase when both $\H_-$ and $\H_+$ are finite dimensional. In that case many analytical difficulties vanish and $L^2(\H_-,\H_+)$ coincides with the set of all linear operators $L(\H_-,\H_+)$. 
In subsequel we will identify $L(\H_-,\H_+)\cong Mat_{M\times N}(\C)\cong \C^{MN}$ with the $M\times N$ complex matrices or with $\C^{MN}$ where $\dim_\C\H_-=M$ and $\dim_\C\H_+=N$.

We define the Poisson bracket on the space of smooth functions $C^\infty(Mat_{M\times N}(\C))$ in the standard way, i.e. 
\be \label{poisson}\{f,g\}(Z,Z^+):=i\Tr\left(\frac{\partial f}{\partial Z}\frac{\partial g}{\partial Z^+}-
\frac{\partial g}{\partial Z}\frac{\partial f}{\partial Z^+}\right)\ee
where $Z\in Mat_{M\times N}(\C)$ and its conjugate $Z^+\in Mat_{N\times M}(\C)$.

In order to obtain a family of Hamiltonians in involution we need to introduce the following notation. Let $\H:=\H_+\oplus\H_-$ and $P_\pm$ be the orthogonal projector on $\H_\pm$.
We will consider a self-adjoint operator $\mu\in L(\H)$ in the block form consistent with the above decomposition of $\H$ as
\be\label{block}\mu:=\left(\begin{matrix}
  A&Z  \\
  Z^+&D 
         \end{matrix}\right)\in L(\H),\ee
where operators $A\in L(\H_+)$ and $D\in L(\H_-)$ are self-adjoint. Moreover we will assume that eigenvalues of both $A$ and $D$ differ from each other. In the following
$A$ and $D$ will be interpreted as parameters of considered Hamiltonian hierarchy.

\begin{prop}
The functions
\be \label{hamiltonians}H_{k,\lambda}:=\Tr (\mu+\lambda P_+)^k,\ee
where $\lambda\in\R$ and $k\in \N$,
are in involution
\be \{H_{k,\lambda}, H_{n,\lambda'}\}=0\ee
with respect to the Poisson bracket \eqref{poisson}.
\end{prop}
\prf{
Without loss of generality 
we can assume that $\lambda'=0$ .
Since the derivative $DH_{k,\lambda}(\mu)$ of $H_{k,\lambda}$ is given by
\be DH_{k,\lambda}(\mu)=k (\mu+\lambda P_+)^{k-1}\ee
we gather that
\be \frac{\partial H_{k,\lambda}}{\partial Z}=ik P_-(\mu+\lambda P_+)^{k-1}P_+.\ee
Thus we obtain
\be\{H_{k,\lambda}, H_{n,0}\}=kn\Tr\big(P_-(\mu+\lambda P_+)^{k-1}P_+\mu^{n-1}P_--\ee
$$-P_-\mu^{n-1}P_+(\mu+\lambda P_+)^{k-1}P_-\big)=$$
$$=ikn\Tr\big((\mu+\lambda P_+)^{k-1}P_+\mu^{n-1}-
\mu^{n-1}P_+(\mu+\lambda P_+)^{k-1}\big)-$$
$$-ikn\Tr\big(P_+(\mu+\lambda P_+)^{k-1}P_+\mu^{n-1}-
P_+\mu^{n-1}P_+(\mu+\lambda P_+)^{k-1}\big).$$
Note that due to the invariance of $\Tr$ under cyclic permutations the second term in the expression above vanishes identically.
Thus we get
\be\{H_{k,\lambda}, H_{n,0}\}=ikn\Tr\big(\mu^{n-1}[(\mu+\lambda P_+)^{k-1},P_+]\big)=\ee
$$=\frac{1}\lambda ikn\Tr\big(\mu^{n-1}[(\mu+\lambda P_+)^{k-1},\mu+\lambda P_+]\big)-$$
$$-\frac{1}\lambda ikn\Tr\big(\mu^{n-1}[(\mu+\lambda P_+)^{k-1},\mu]\big)=$$
$$=-\frac{1}\lambda ikn \Tr\big((\mu+\lambda P_+)^{k-1}[\mu,\mu^{n-1}]\big)=0.
$$
}

Let us note here that functions $H_{k,\lambda}$ defined in \eqref{hamiltonians} 
form an integrable hamiltonian hierarchy. 
This hierarchy possesses two extra families of integrals of motion, which are analogues of Manley--Rowe integrals, see e.g. \cite{holm1}.

\begin{prop}
The functions 
\be \label{a-d}\alpha_k:=\Tr(A^kZZ^+)\qquad \delta_k:=\Tr(D^k Z^+Z),\ee 
$k\in\N$, commute with each other and with Hamiltonians \eqref{hamiltonians}
\be \{\alpha_k,\alpha_l\}=\{\alpha_k,\delta_l\}=\{\delta_k,\delta_l\}=0.\ee

\end{prop}
\prf{
We note that
\be \frac{\partial\alpha_k}{\partial Z}=Z^+A^k\qquad
\frac{\partial\alpha_k}{\partial Z^+}=A^kZ\ee
\be \frac{\partial\delta_k}{\partial Z}=D^kZ^+\qquad
\frac{\partial\delta_k}{\partial Z^+}=ZD^k.\ee
Thus we have
\be \{\alpha_k,\alpha_l\}=i\Tr(Z^+A^k A^lZ - A^kZ Z^+A^l)=0\ee
\be \{\delta_k,\delta_l\}=i\Tr(D^kZ^+ZD^l - ZD^k D^lZ^+)=0\ee
\be \{\alpha_k,\delta_l\}=i\Tr(Z^+A^k ZD^l - A^kZ D^l Z^+)=0.\ee
Moreover we have 
\be \label{p23}\{\alpha_k,H_{l,\lambda}\}=
il\Tr\big(Z^+A^k P_+(\mu+\lambda P_+)^{l-1} P_-A^kZP_-(\mu+\lambda P_+)^{l-1}P_+\big)=\ee
$$=il\Tr\big(P_-\mu (P_+\mu P_+)^k (\mu+\lambda P_+)^{l-1} - (P_+\mu P_+)^k\mu P_-(\mu+\lambda P_+)^{l-1}\big)=$$
$$=
il\Tr\big(\mu (P_+\mu P_+)^k (\mu+\lambda P_+)^{l-1} - (P_+\mu P_+)^k\mu(\mu+\lambda P_+)^{l-1}\big)-$$
$$-il\Tr\big(P_+\mu (P_+\mu P_+)^k (\mu+\lambda P_+)^{l-1} - (P_+\mu P_+)^k\mu P_+(\mu+\lambda P_+)^{l-1}\big)=
$$
$$=il\Tr\big(\mu[(P_+\mu P_+)^k, (\mu+\lambda P_+)^{l-1}]\big)-
il\Tr\big((P_+\mu P_+)^{k+1} (\mu+\lambda P_+)^{l-1} -$$
$$- (P_+\mu P_+)^{k+1}(\mu+\lambda P_+)^{l-1}\big)=il\Tr\big((\mu+\lambda P_+)[(P_+\mu P_+)^k, (\mu+\lambda P_+)^{l-1}]\big)-$$
$$-il\lambda \Tr\big(P_+[(P_+\mu P_+)^k, (\mu+\lambda P_+)^{l-1}]\big)=0.$$
Proof that $\{\delta_k,H_{l,\lambda}\}=0$ is analogous to \eqref{p23}.
}

The Hamilton equations generated by any $H\in C^\infty(Mat_{M\times N}(\C))$ given by the Poisson bracket \eqref{poisson} are the following
\be \label{eq-Z1}\dot Z = i\frac{\partial H}{\partial Z^+}\qquad
\dot Z^+ = -i\frac{\partial H}{\partial Z}.\ee

In the case when $H=H_{k,\lambda}$ they assume the form
\begin{align} \label{eq-Z}\dot Z &= kP_+(\mu+\lambda P_+)^{k-1}P_-\\
\nonumber \dot Z^+ &=kP_-(\mu+\lambda P_+)^{k-1}P_+.\end{align}
These equations are in general nonlinear. Even if the family of integrals of motion in involution is big enough, it may be technically difficult to find their explicit solutions. Thus in next sections we will restrict our considerations to several specific cases when it is possible to solve the system in the explicit way.

Combining the Hamiltonians \eqref{hamiltonians} for $\lambda=0$ and $k=4,5$
\begin{align}
 H_{4,0}&=\Tr \mu^4 = \Tr \big(A^4 + D^4+
4D^2Z^+Z+
4AZDZ^++\\
&+4A^2ZZ^++
2(Z^+Z)^2\big)\nonumber \\
\label{int5} 
H_{5,0}&=\Tr\mu^5=\Tr\big(A^5+D^5+
5D^2Z^+AZ+
5DZ^+A^2Z+
\\
&+5A(ZZ^+)^2+5D(Z^+Z)^2+
5A^3ZZ^++
5D^3 Z^+Z\big)\nonumber
 \end{align}
with Hamiltonians $\alpha_k$ and $\delta_k$ for $k=1,2,3$ we find that the two following functions 
\begin{align}\label{hhhh}
H&=\frac12\Tr(Z^+Z)^2+\Tr(AZDZ^+)\\ 
F&=\Tr A(ZZ^+)^2+\Tr(D^2Z^+AZ+DZ^+A^2Z)
\end{align}
belong to the hierarchy of Hamiltonians generated by \eqref{hamiltonians} and \eqref{a-d}. 

Equation of motion with respect to Hamiltonian \eqref{hhhh} are the following
\begin{align} \label{eq-Z2}\dot Z &= AZD+ZZ^+Z\\
\nonumber \dot Z^+ &=DZ^+A+Z^+ZZ^+.
\end{align}
They can be considered as a pair of coupled Ricatti-type equations on the variables $Z$ and $Z^+$.

\section{Solution in $(2+3)$-dimensional case}\label{sec:2+3}

In this section we consider in details the Hamilton equations in the case when $\dim \H_+=2$ and $\dim \H_-=3$.
Without loss of generality we can assume that the matrices $A$ and $D$ are diagonal. So the matrix $\mu$ defined in \eqref{block} assumes the form
\be \mu=\begin{pmatrix}a_1 & 0 & z_{1} & z_{2} & z_{3}\cr 0 & a_2 & v_{1} & v_{2} & v_{3}\cr \bar z_{1} & \bar v_{1} & d_1 & 0 & 0\cr \bar z_{2} & \bar v_{2} & 0 & d_2 & 0\cr \bar z_{3} & \bar v_{3} & 0 & 0 & d_3\end{pmatrix},\ee
where $z_k,v_k\in \C$ and $a_k, d_k\in\R$.

Expressing the integrals of motion $H$, $F$, $\alpha_0$, $\alpha_1$, $\delta_1$, $\delta_2$ in the vector coordinates $z=(z_1,z_2,z_3)^T$ and $v=(v_1,v_2,v_3)^T$
we obtain the following six functionally independent integrals of motion in involution:
\begin{align}\label{H1}
H&=\frac12 (z^+z)^2+\frac12 (v^+v)^2+\abs{v^+z}^2+a_1z^+Dz+a_2v^+Dv\\
F&=a_1(z^+z)^2+a_2 (v^+v)^2+(a_1+a_2)\abs{v^+z}^2+z^+z z^+Dz+v^+v v^+Dv
+\nonumber \\
&+v^+z z^+Dv+ z^+v v^+Dz+a_1^2z^+Dz+a_2^2v^+Dv+a_1 z^+D^2z+a_2v^+D^2v\\
\alpha_0&=\delta_0=z^+z+v^+v\\
\alpha_1&=a_1 z^+z+a_2v^+v\\
\delta_1&=z^+Dz+v^+Dv\\
\delta_2&=z^+D^2z+v^+D^2v.\label{H6}
\end{align}

Now we define for $k=1,2,3$ functions
\be\label{f-u2}\eta_k:=\bar{v_k}z_k,\qquad r_k:=\abs{z_k}^2-\abs{v_k}^2, \qquad s_k:=\abs{z_k}^2+\abs{v_k}^2\ee
on the phase space $\C^6$. They span the Lie algebra with respect to the Poisson bracket \eqref{poisson}:
\begin{align} \label{alg-pb}\{\eta_k,\bar\eta_l\}&=i r_k\delta_{kl}\nonumber\\
\{r_k,\eta_l\}&=2i\eta_k\delta_{kl}\nonumber\\
\{r_k,\bar\eta_l\}&=-2i\bar\eta_k\delta_{kl}\\
\{s_k,s_l\}&=\{s_k,r_l\}=\{s_k,\eta_l\}=\{s_k,\bar\eta_l\}=0\nonumber\\
\{\eta_k,\eta_l\}&=\{\bar\eta_k,\bar\eta_l\}=0\nonumber
\end{align}
where $k,l=1,2,3$.

Let us consider Lie--Poisson space $\u(2)^*$ dual to the Lie algebra 
$$\u(2):=\{X\in Mat_{2\times2}(\C)\;|\; X^++X=0\}$$ 
of the unitary group $U(2)$. We identify $\u(2)^*$ with $\u(2)$ by the pairing 
\be \pair{X}{Y}:=\Tr(XY)\ee
and introduce the following coordinates
\be \label{coord-u2} X=i\begin{pmatrix} \frac{s+r}2 & \eta\\
\bar\eta & \frac{s-r}2\end{pmatrix}\ee
on $\u(2)^*$.
In these coordinates the Lie--Poisson bracket 
\be \{f,g\}_{LP}(X):=\pair{X}{[Df(X),Dg(X)]}\ee 
assumes the following form
\be \label{u2-pb}\{f,g\}_{LP}(s,r,\eta,\bar\eta)=i\left(r\left(\frac{\partial f}{\partial \eta}
\frac{\partial g}{\partial \bar\eta}-\frac{\partial f}{\partial \bar\eta}\frac{\partial g}{\partial \eta}\right)\right.+\ee
$$\left.+2\eta\left(\frac{\partial f}{\partial r}\frac{\partial g}{\partial \eta}-\frac{\partial f}{\partial \eta}\frac{\partial g}{\partial r}\right)
+2\bar\eta\left(\frac{\partial f}{\partial \bar\eta}\frac{\partial g}{\partial r}-
\frac{\partial f}{\partial r}\frac{\partial g}{\partial \bar\eta}\right)\right).$$
Note that functions $s_1$, $s_2$, $s_3$, and 
\be c(r,\eta,\bar\eta):=\frac{r^2}2+2\abs{\eta}^2\ee
are Casimirs for the Lie--Poisson bracket \eqref{u2-pb}.

As it follows from \eqref{alg-pb} the Lie algebra generated by functions \eqref{f-u2} is isomorphic to direct sum $\u(2)\oplus \u(2)\oplus \u(2)$ of three copies of $\u(2)$. The map
\be J(z^+,z,v^+,v):=\bigoplus_{k=1}^3
i\begin{pmatrix} \abs{z_k}^2 & \bar v_k z_k\\
v_k\bar z_k & \abs{v_k}^2\end{pmatrix}\ee
is a Poisson map (momentum map) of $(\C^6,\pb)$ into 
$(\u(2)^*\oplus \u(2)^*\oplus \u(2)^*,\pb_{LP})$, i.e.
\be \{f\circ J,g\circ J\}=\{f,g\}_{LP}\circ J,\ee
where now by $\pb_{LP}$ we denote sum of three copies of Lie--Poisson bracket \eqref{u2-pb}.


Writting the Hamiltonians \eqref{H1}-\eqref{H6} in terms of the variables \eqref{f-u2} we obtain on 
$\u(2)^*\oplus \u(2)^*\oplus \u(2)^*$ the following three Hamiltonians in involution
\begin{align}\label{HH1}
R&:=\alpha_0-\frac2{a_2-a_1}\alpha_1=r_1+r_2+r_3\end{align}
\begin{align}
H&=(\eta_1+\eta_2+\eta_3)(\bar\eta_1+\bar\eta_2+\bar\eta_3)+\frac14(r_1+r_2+r_3)^2+\\
\nonumber&+\frac12(a_1-a_2)(d_1r_1+d_2r_2+d_3r_3)+\frac14(s_1+s_2+s_3)^2+\\
\nonumber&+\frac12(a_1+a_2)(d_1s_1+d_2s_2+d_3s_3)
\end{align}
\begin{align}
G&:= F-(a_1+a_2)H+\frac12(a_2-a_1)(s_1+s_2+s_3)R+a_1a_2\delta_1-\label{HH3}\\
\nonumber&-\frac12(a_1+a_2)(d_1^2s_1+d_2^2s_2+d_3^2s_3)-\\
\nonumber&-\frac12(s_1+s_2+s_3)(d_1s_1+d_2s_2+d_3s_3)=\\
\nonumber&=(\eta_1+\eta_2+\eta_3)(d_1\bar\eta_1+d_2\bar\eta_2+d_3\bar\eta_3)+\\
\nonumber&+(\bar\eta_1+\bar\eta_2+\bar\eta_3)(d_1\eta_1+d_2\eta_2+d_3\eta_3)+\\
\nonumber&+\frac12(a_1-a_2)(d_1^2r_1+d_2^2r_2+d_3^2r_3)+\\
\nonumber&+\frac12(r_1+r_2+r_3)(d_1r_1+d_2r_2+d_3r_3)\\
\end{align}
and the Casimirs
\begin{align}\label{C1}
\delta_0&=s_1+s_2+s_3\\
\delta_1&=d_1s_1+d_2s_2+d_3s_3\\
\delta_2&=d_1^2s_1+d_2^2s_2+d_3^2s_3\label{C3}
\end{align}
which are expressed by the Casimirs $s_1$, $s_2$, and $s_3$.

Since one has 
\be \label{cas}\frac12 c_k=s_k^2=r_k^2+(2\abs{\eta_k})^2\ee
for $k=1,2,3$ we conclude from \eqref{C1}-\eqref{C3} that if $s_k\neq 0$ then the Hamiltonian hierarchy \eqref{H1}-\eqref{H6} is reduced to the three Hamiltonians \eqref{HH1}-\eqref{HH3} defined on the six-dimensional symplectic leaves 
\be \label{sympl-leaf}\Sigma_{s_1,s_2,s_3}:=S^2_{s_1}\times S^2_{s_2}\times S^2_{s_3}\ee
which are the products of two-dimensional spheres with radii $s_1$, $s_2$, and $s_3$ respectively.

In order to simplify notation we will use the polar coordinates $r_k,\phi_k$ on $\Sigma_{s_1,s_2,s_3}$ defined by
\be \eta_k=\frac12\sqrt{s_k^2-r_k^2}\;e^{i2\phi_k}.\ee
Note here that $r_k$ and $\phi_k$, $-s_k\leq r_k\leq s_k$ and $-\frac\pi2\leq\phi_k\leq\frac\pi2$, where $k=1,2,3$ and $s_k>0$, form canonical system of coordinates (Darboux coordinates) on 
$\Sigma_{s_1,s_2,s_3}$, i.e.
\be \label{pb-rk}\{r_k,\phi_l\}=\delta_{kl}.\ee
From \eqref{pb-rk} it follows that the symplectic form on $\Sigma_{s_1,s_2,s_3}$ is given by 
\be\omega_{s_1,s_2,s_3}=dr_1\wedge d\phi_1+dr_2\wedge d\phi_2+dr_3\wedge d\phi_3.\ee 
In the polar coordinates the Hamiltonians \eqref{HH1}-\eqref{HH3} take the following form
\begin{align}
H\label{H-phi}&=\frac12\sqrt{(s_1^2-r_1^2)(s_2^2-r_2^2)}\cos(\phi_1-\phi_2)+\\
\nonumber&+\frac12\sqrt{(s_1^2-r_1^2)(s_3^2-r_3^2)}\cos(\phi_1-\phi_3)\\
\nonumber&+\frac12\sqrt{(s_2^2-r_2^2)(s_3^2-r_3^2)}\cos(\phi_2-\phi_3)
-\frac14(r_1^2+r_2^2+r_3^2)+\\
\nonumber&+\frac14(s_1^2+s_2^2+s_3^2)
+\frac14(r_1+r_2+r_3)^2+\\
\nonumber&+\frac12(a_1-a_2)(d_1r_1+d_2r_2+d_3r_3)+\frac14(s_1+s_2+s_3)^2+\\
\nonumber&+\frac12(a_1+a_2)(d_1s_1+d_2s_2+d_3s_3)\end{align}
\begin{align}
G\label{G-phi}&= 
\frac12(d_1+d_2)\sqrt{(s_1^2-r_1^2)(s_2^2-r_2^2)}\cos(\phi_1-\phi_2)+\\
\nonumber&+\frac12(d_1+d_3)\sqrt{(s_1^2-r_1^2)(s_3^2-r_3^2)}\cos(\phi_1-\phi_3)+\\
\nonumber&+\frac12(d_2+d_3)\sqrt{(s_2^2-r_2^2)(s_3^2-r_3^2)}\cos(\phi_2-\phi_3)-\\
\nonumber&-\frac12(d_1r_1^2+d_2r_2^2+d_3r_3^2)+
\frac12(d_1s_1^2+d_2s_2^2+d_3s_3^2)+\\
\nonumber&+\frac12(a_1-a_2)(d_1^2r_1+d_2^2r_2+d_3^2r_3)+\\
\nonumber&+\frac12(r_1+r_2+r_3)(d_1r_1+d_2r_2+d_3r_3)\end{align}
\begin{align}
R&=r_1+r_2+r_3.\label{HHH3}
\end{align}
The function $R$ generates on the symplectic leave $\Sigma_{s_1,s_2,s_3}$ the following Hamiltonian flow 
\be \label{sigma3}\sigma^R_\phi(r_1,r_2,r_3,\phi_1,\phi_2,\phi_3)=
(r_1,r_2,r_3,\phi_1+\phi,\phi_2+\phi,\phi_3+\phi).\ee
Reducing the Hamiltonian system $(\Sigma_{s_1,s_2,s_3},\omega_{s_1,s_2,s_3},H,G,R)$ to the level $R^{-1}(r)$ of the function $R$ defined in \eqref{HHH3} we obtain 
the reduced phase space  $R^{-1}(r)/\{\sigma^R_\phi\}_{\phi\in\R}$ with symplectic form $\omega$ given by 
\be \omega=dr_1\wedge d\psi_1+dr_2\wedge d\psi_2\ee
where $\psi_1:=\phi_1-\phi_3$ and $\psi_2:=\phi_2-\phi_3$.

By introducing the notation
\begin{align}
f(r_1,r_2)&:=\frac12\sqrt{(s_1^2-r_1^2)(s_2^2-r_2^2)}\\
f_1(r_1,r_2)&:=\frac12\sqrt{(s_1^2-r_1^2)(s_3^2-(r-r_1-r_2)^2)}\\
f_2(r_1,r_2)&:=\frac12\sqrt{(s_2^2-r_2^2)(s_3^2-(r-r_1-r_2)^2)}\\
h(r_1,r_2)&:=-\frac14(r_1^2+r_2^2+(r-r_1-r_2)^2)+\\
\nonumber&+\frac14(s_1^2+s_2^2+s_3^2)
+\frac14r^2+\\
\nonumber&+\frac12(a_1-a_2)(d_1r_1+d_2r_2+d_3(r-r_1-r_2))+\\
\nonumber&+\frac14(s_1+s_2+s_3)^2+\frac12(a_1+a_2)(d_1s_1+d_2s_2+d_3s_3)\\
g(r_1,r_2)&:=-\frac12(d_1r_1^2+d_2r_2^2+d_3(r-r_1-r_2)^2)+\\
\nonumber&+\frac12(d_1s_1^2+d_2s_2^2+d_3s_3^2)+\\
\nonumber&+\frac12(a_1-a_2)(d_1^2r_1+d_2^2r_2+d_3^2(r-r_1-r_2))+\\
\nonumber&+\frac r2(d_1r_1+d_2r_2+d_3(r-r_1-r_2))
\end{align}
we can write the Hamiltonians \eqref{H-phi} and \eqref{G-phi} as follows
\begin{align}
\label{H-psi}
\nonumber H&=f(r_1,r_2)\cos(\psi_1-\psi_2)+f_1(r_1,r_2)\cos(\psi_1)+\\
&+f_2(r_1,r_2)\cos(\psi_2)+h(r_1,r_2)\\
\nonumber\label{G-psi}G&=(d_1+d_2)f(r_1,r_2)\cos(\psi_1-\psi_2)+(d_1+d_3)f_1(r_1,r_2)\cos(\psi_1)+\\
&+(d_2+d_3)f_2(r_1,r_2)\cos(\psi_2)+g(r_1,r_2).
\end{align}


From the above formulas we find that the Hamilton equations for the Hamiltonian \eqref{H-psi} in the canonical coordinates $r_1$, $r_2$, $\psi_1$, $\psi_2$ assume the following form
\begin{align}
\label{eq-psi1} \dot \psi_1&=\{\psi_1,H\}= -\frac{\partial f}{\partial r_1}\cos(\psi_1-\psi_2)-\frac{\partial f_1}{\partial r_1}\cos(\psi_1)-\\
\nonumber&-\frac{\partial f_2}{r_1}\cos(\psi_2)-\frac{\partial h}{\partial r_1}\\
\label{eq-psi2}\dot \psi_2&=\{\psi_2,H\}= -\frac{\partial f}{\partial r_2}\cos(\psi_1-\psi_2)-\frac{\partial f_1}{\partial r_2}\cos(\psi_1)-\\
\nonumber&-\frac{\partial f_2}{\partial r_2}\cos(\psi_2)-\frac{\partial h}{\partial r_2}\\
\label{eq-r1}\dot r_1&=\{r_1,H\}=-f(r_1,r_2)\sin(\psi_1-\psi_2)-f_1(r_1,r_2)\sin(\psi_1)\\
\label{eq-r2}\dot r_2&=\{r_2,H\}=f(r_1,r_2)\sin(\psi_1-\psi_2)-f_2(r_1,r_2)\sin(\psi_2)
\end{align}


Equations \eqref{H-psi}-\eqref{G-psi} define 
$\psi_1$ and $\psi_2$ as an implicit function of $r_1$ and $r_2$. 
In order to solve equations \eqref{eq-psi1}-\eqref{eq-r2} we apply the generating function method. Thus we introduce new variables $\gamma$ and $\tau$ canonically conjugated to the integrals of motion $G$ and $H$, i.e.
\be \label{fs}\omega=-d(\psi_1 dr_1+\psi_2 dr_2)=d(\gamma dG+\tau dH).\ee
From \eqref{fs} it follows that there exists a locally defined function $\Phi(r_1,r_2,G,H)$
such that
\be \label{dphi}d\Phi=\psi_1 dr_1+\psi_2dr_2+\gamma dG+\tau dH.\ee
The equality \eqref{dphi} gives the relationship between new canonical coordinates $(\gamma,\tau,G,H)$ and the old canonical coordinates $(r_1,r_2,\psi_1,\psi_2)$ given that we obtain the generating function $\Phi$.
To this end let us note that consistency condition 
\be\frac{\partial\psi_1}{\partial r_2}=\frac{\partial\psi_2}{\partial r_1}\ee
for the equations 
\begin{align}
\label{Phi-psi1}\frac{\partial\Phi}{\partial r_1}&=\psi_1\\
\label{Phi-psi2}\frac{\partial\Phi}{\partial r_2}&=\psi_2
\end{align}
follows from $\{H,G\}=0$. The proof of this fact is a consequence of the formula on the derivative of the implicit function
\be{\Large\begin{pmatrix}
  \frac{\partial \psi_1}{\partial r_1}& \frac{\partial \psi_1}{\partial r_2} \\
  \frac{\partial \psi_2}{\partial r_1}& \frac{\partial \psi_2}{\partial r_2}
\end{pmatrix}=-
\begin{pmatrix}
       \frac{\partial G}{\partial \psi_1}& \frac{\partial G}{\partial \psi_2}\\
	\frac{\partial H}{\partial \psi_1} & \frac{\partial H}{\partial \psi_2}
      \end{pmatrix}^{-1}
\begin{pmatrix}
  \frac{\partial G}{\partial r_1}& \frac{\partial G}{\partial r_2} \\
  \frac{\partial H}{\partial r_1}& \frac{\partial H}{\partial r_2}
\end{pmatrix}.}
\ee

Thus we can define $\Phi$ for $(r_1,r_2)\in[-s_1,s_1]\times[-s_2,s_2]$ by
\be \label{Phi-int}\Phi(r_1,r_2,G,H)=-\int_0^1\big(\psi_1(s r_1,s r_2)\,r_1+\psi_2(s r_1,s r_2)\,r_2\big)ds.\ee
Recall that the dependence of $\psi_1$ and $\psi_2$ on $r_1$, $r_2$, $G$, $H$ is given in the implicit way by \eqref{H-psi}-\eqref{G-psi}.

Now submitting $\Phi$ defined by \eqref{Phi-int} into
\begin{align}
\label{Phi_G}\frac{\partial\Phi}{\partial G}&=\gamma \\
\label{Phi_H}\frac{\partial\Phi}{\partial H}&=\tau
\end{align}
we find the new coordinates $\gamma$ and $\tau$.

Since the variables $(\gamma,\tau,G,H)$ are angle--action coordinates for the Hamiltonian system defined by $H$ we obtain
\begin{align}
 \tau(t) &= t+t_0\\
 \gamma(t) &= \const\\
 G(t)&= \const\\
  H(t)&=\const
\end{align}

The time dependence of the variables $r_1(t)$ and $r_2(t)$ can be found in the implicit way by \eqref{Phi_G}-\eqref{Phi_H}. Next we obtain the time dependence of $\psi_1(t)$ and $\psi_2(t)$ from \eqref{Phi-psi1}-\eqref{Phi-psi2}.

Summing up, we have solved Hamiltonian system given by Hamiltonian \eqref{H1} describing nonlinear interaction of six waves. The solution was obtained in quadratures, but due to technical difficulties, explicit form of solution would be too complicated to present.

\section{Solution in $(2+2)$-dimensional case}\label{sec:2+2}

In this section we consider a special situation of $(2+3)$-dimensional case, when $s_3=0$. From \eqref{cas} we see that $s_3=0$ implies that $r_3=0$ and $\abs{\eta_3}=0$. 
Since $s_3$ is Casimir, this anzatz is consistent with the evolution with respect to all Hamiltonians \eqref{hamiltonians}.

Therefore we will solve equation \eqref{eq-Z2} in the $(2+2)$-dimensional case taking instead of 6-dimensional symplectic leaves \eqref{sympl-leaf}, the 4-dimensional symplectic leaves
\be \label{sympl-leaf4}\Sigma_{s_1,s_2}:=S^2_{s_1}\times S^2_{s_2}.\ee
with the symplectic form given by 
\be\omega_{s_1,s_2}=dr_1\wedge d\phi_1+dr_2\wedge d\phi_2.\ee 
The Hamiltonians \eqref{HH1}-\eqref{HH3} assume now the form
\begin{align}
H&=\frac12\sqrt{(s_1^2-r_1^2)(s_2^2-r_2^2)}\cos(\phi_1-\phi_2)-\frac14(r_1^2+r_2^2)+\\
\nonumber&+\frac14(s_1^2+s_2^2)
+\frac14(r_1+r_2)^2+\frac12(a_1-a_2)(d_1r_1+d_2r_2)+\\
\nonumber&+\frac14(s_1+s_2)^2+\frac12(a_1+a_2)(d_1s_1+d_2s_2)\\
G&= 
\frac12(d_1+d_2)\sqrt{(s_1^2-r_1^2)(s_2^2-r_2^2)}\cos(\phi_1-\phi_2)-\\
\nonumber&-\frac12(d_1r_1^2+d_2r_2^2)+
\frac12(d_1s_1^2+d_2s_2^2)+
\frac12(a_1-a_2)(d_1^2r_1+d_2^2r_2)+\\
\nonumber&+\frac12(r_1+r_2)(d_1r_1+d_2r_2)\\
R&=r_1+r_2.
\end{align}

Note that in this case no longer we need $G$ as an additional integral of motion, since in $(2+2)$-dimensional case the integrals of motion $H$ and $R$ are sufficient to integrate the system. Similarly to the $(2+3)$-dimensional case, we reduce the Hamiltonian system $(\Sigma_{s_1,s_2},\omega_{s_1,s_2},H,R)$ to the level set $R^{-1}(r)$ and obtain the Hamiltonian on the reduced phase space $R^{-1}(r)/\{\sigma^R_\phi\}_{\phi\in\R}$
\be H\label{H2-cos}=\frac12\sqrt{(s_1^2-r_1^2)(s_2^2-(r-r_1)^2)}\cos(\psi_1)+w_2(r_1),\ee
where
\begin{align}
w_2(r_1)&:=-\frac14(r_1^2+(r-r_1)^2)+\frac14(s_1^2+s_2^2)
+\\
\nonumber&+\frac14 r^2+\frac12(a_1-a_2)(d_1r_1+d_2(r-r_1))+\\
\nonumber&+\frac14(s_1+s_2)^2+\frac12(a_1+a_2)(d_1s_1+d_2s_2).
\end{align}
The symplectic form $\omega$ on $R^{-1}(r)/\{\sigma^R_\phi\}_{\phi\in\R}$ is given by 
\be \omega=dr_1\wedge d\psi_1,\ee
where $\psi_1:=\phi_1-\phi_2$. The flow $\{\sigma_\phi^R\}_{\phi\in\R}$ defined by the integral of motion $R$ has the form
\be \sigma^R_\phi(r_1,r_2,\phi_1,\phi_2)=(r_1,r_2,\phi_1+\phi,\phi_2+\phi).\ee

The equations of motion \eqref{eq-psi1}-\eqref{eq-r2} in this case reduces the following ones
\begin{align}
\label{phi1_2}\dot \psi_1&=\{\psi_1,H\}=r_1- \frac12 r+\frac12(a_1+a_2)(d_1-d_3)-\\
\nonumber&-\frac{-r_1(s_2^2-(r-r_1)^2)+(s_1^2-r_1^2)(r-r_1)}{2\sqrt{(s_1^2-r_1^2)(s_2^2-(r-r_1)^2)}}\cos\psi_1\\
\label{r1_2}\dot r_1&=\{r_1,H\}=-\frac12\sqrt{(s_1^2-r_1^2)(s_2^2-(r-r_1)^2)}\sin\psi_1.
\end{align}
In order to solve the equation \eqref{r1_2} we use \eqref{H2-cos} and Pythagorean identity
to 
obtain the relation
\be \label{r1x}4(H-w(r_1))^2+4(\dot r_1)^2=(s_1^2-r_1^2)(s_2^2-(r-r_1)^2).\ee
The solution of \eqref{r1x} is in the form of elliptic integral of the first kind
\be t=\pm\int\frac{r_1\;dr_1}{\sqrt{w_4(r_1)}},\ee
where 
\be w_4(r_1):=(s_1^2-r_1^2)(s_2^2-(r-r_1)^2)-4(H-w_2(r_1))^2\ee
is a polynomial of fourth order. Thus $r_1$ is an elliptic function of the parameter $t$. Subsequently, from \eqref{H2-cos} we obtain
\be \psi_1=\arccos\frac{2(H-w_2(r_1))}{\sqrt{(s_1^2-r_1^2)(s_2^2-(r-r_1)^2)}}\ee

In the analogous way we can find solution of the Hamiltonian system given by $G$ or, more generally, for any Hamiltonian $H_{4,\lambda}$ or $H_{5,\lambda}$.

\section{Physical interpretation of the (2+3)-mode Hamiltonian}\label{sec:phys}

In order to elucidate the optical interpretation of the Hamiltonian \eqref{H1} let us express it in the coordinates $z_1$, $z_2$, $z_3$ and $v_1$, $v_2$, $v_3$. 
\begin{align}\label{H1-modes}
H&= a_1d_1\abs{z_1}^2+a_1d_2\abs{z_2}^2+a_1d_3\abs{z_3}^2+a_2d_1\abs{v_1}^2+\\
\nonumber&+a_2d_2\abs{v_2}^2+a_2d_3\abs{v_3}^2+\\
\nonumber&+\frac12(\abs{z_1}^4+\abs{z_2}^4+\abs{z_3}^4+\abs{v_1}^4+\abs{v_2}^4+\abs{v_3}^4)+\\
\nonumber&+\abs{z_1}^2\abs{z_2}^2+\abs{z_1}^2\abs{z_3}^2+\abs{z_2}^2\abs{z_3}^2+\abs{v_1}^2\abs{v_2}^2+\abs{v_1}^2\abs{v_3}^2+\abs{v_2}^2\abs{v_3}^2+\\
\nonumber&+\abs{z_1}^2\abs{v_1}^2+\abs{z_2}^2\abs{v_2}^2+\abs{z_3}^2\abs{v_3}^2+\\
\nonumber&+\bar v_1v_2 \bar z_2 z_1+\bar v_2v_1 \bar z_1 z_2+\bar v_1v_3 \bar z_3 z_1+\bar v_3v_1 \bar z_1 z_3+\bar v_2v_3 \bar z_3 z_2+\bar v_3v_2 \bar z_2 z_3,
\end{align}
Assuming that these coordinates describe separate modes of the six-wave interacting nonlinearly through the nonlinear dielectric  medium, we get the following optical interpretation of the particular terms:
%
\ben[i)]
	\item the quadratic terms $\abs{z_i}^2$ and $\abs{v_i}^2$ constitute free Hamiltonian $H_0$, i.e. they describe free energy of the light, where $a_id_j$ is proportional to the corresponding modes frequency;
	\item the terms $\abs{z_i}^4$ and $\abs{v_i}^4$ are responsible for the Kerr effect, i.e. third order of nonlinearity of polarisation of a medium causing intensity-dependent phase shift, see e.g. \cite{walls-milburn};
	\item the terms $\abs{z_i}^2\abs{z_j}^2$, $\abs{z_i}^2\abs{v_j}^2$, $\abs{v_i}^2\abs{v_j}^2$ introduce Kerr-like effect, i.e. phase shift of $i^{th}$ mode depending on intensity of $j^{th}$ mode;
	\item the other terms describe the conversion between the modes, e.g. the term 
$\bar v_1v_2 \bar z_2 z_1$ describes the process of absorption by the medium of certain amount of light in modes $z_1$ and $v_2$ with simultaneous emission of light in modes $v_1$ and $z_2$.
\een


\end{document}